**Title**

# Observation of Inverse Edelstein Effect in Rashba-Split 2DEG between $SrTiO_3$ and $LaAlO_3$ at Room Temperature

**Authors**

Qi Song[1,2†], Hongrui Zhang[3†], Tang Su[1,2], Wei Yuan[1,2], Yangyang Chen[1,2], Wenyu Xing[1,2], Jing Shi[4*], Ji Rong Sun[3*], and Wei Han[1,2*]

**Affiliations**

[1]International Center for Quantum Materials, School of Physics, Peking University, Beijing 100871, China.

[2]Collaborative Innovation Center of Quantum Matter, Beijing 100871, China.

[3]Beijing National Laboratory for Condensed Matter Physics and the Institute of Physics, Chinese Academy of Sciences, Beijing 100190, China.

[4]Department of Physics and Astronomy, University of California, Riverside, California 92521, USA.

†These authors contributed equally to the work

*Correspondence to: weihan@pku.edu.cn (W.H.); jrsun@iphy.ac.cn (J.R.S.); and jing.shi@ucr.edu (J.S.)

## Abstract

The Rashba physics has been intensively studied in the field of spin orbitronics, for the purpose of searching novel physical properties and the FM magnetization switching for technology applications. Here, we report the observation of the inverse Edelstein effect up to room temperature in the Rashba-split two dimensional electron gas (2DEG) between two insulating oxides $SrTiO_3$ and $LaAlO_3$ with the $LaAlO_3$ layer thickness from 3 to 40 unit cells (UC). We further demonstrate that the spin voltage could be dramatically manipulated by electric field effect for the 2DEG between $SrTiO_3$ and 3 UC $LaAlO_3$. These results demonstrate that the Rashba-split 2DEG at the complex oxide interface can be used for room temperature efficient charge-and-spin conversion for the generation and detection of spin current.

**One sentence summary:** Gate tunable inverse Edelstein effect is observed at room temperature in Rashba-split 2DEG at the complex oxides interface.

## Introduction

In 1990, Edelstein predicted that spin current could be induced by charge current flowing in inversion asymmetric two dimensional electron gases (2DEG), which is often referred as the Edelstein effect (EE) (*1*). The magnitude of EE highly depends on the Rashba spin orbit coupling, which provides a locking between the momentum and spin polarization directions (*2*). The generated spin current density could be described in the following form:

$$J_S \propto \alpha_R (\hbar/e)(\vec{z} \times \vec{j}_C) \quad (1)$$

where $\alpha_R$ is the Rashba parameter, $\vec{z}$ is the interfacial electric field direction perpendicular to the 2DEG, and $\vec{j}_C$ is the charge current. The opposite of EE is often called inverse Edelstein effect (IEE), which means that spin accumulation in inversion asymmetric 2DEG could generate an in-plane electric field perpendicular to the spin polarization direction (*3*). Due to the potential highly efficient spin-and-charge conversion, both the EE and IEE have attracted a great deal of interest for spintronics and various experiments have been performed on the Rashba interfaces between two metallic films (*4-7*), two-dimensional materials (*8-12*), and the topological surface states (*13-18*).

Here, we report the observation of the IEE in Rashba-split 2DEG between two insulating oxides $SrTiO_3$ and $LaAlO_3$ up to room temperature. The spin current in the 2DEG is generated by spin pumping from a ferromagnetic $Ni_{80}Fe_{20}$ (Py) electrode through $LaAlO_3$ layer of the thickness up to 40 unit cells (UCs). The IEE is probed by measuring the electric voltage that is created by the spin-to-charge conversion of the injected spin current due to the Rashba spin orbit coupling of the 2DEG. Furthermore, we demonstrate that the spin voltage in the Rashba-split 2DEG between $SrTiO_3$ and 3 UC $LaAlO_3$ could be switched on/ off using a perpendicular electric field. These results show that the complex oxide interface, that has proven to show many interesting physical properties (*19-23*), can be used for efficient charge-and-spin conversion, which is gate controllable, for the spin current generation, spin detection, and the manipulation of the magnetization.

## Results

The Rashba-split 2DEG is formed between the (001)-oriented $SrTiO_3$ and $LaAlO_3$, as shown in Fig. 1A (*24, 25*). The $LaAlO_3$ layers from 3 to 40 UCs are grown on $SrTiO_3$ substrates via pulsed laser deposition (see Methods for details) and the thickness of the $LaAlO_3$ is monitored by *in situ* reflective high energy electron diffraction (RHEED) oscillations, as shown in Fig 1B. Fig. 1C illustrates the expected qualitative energy diagram of the Rashba-split 2DEG thus obtained. At the Fermi level ($E_F$), the spin textures of the outer-circle and the inner-circle are opposite to each other. Whether the spin texture is clockwise or counter clockwise depends on the sign of $\alpha_R$ and the definition of the normal direction (*26*). The $\alpha_R$ for the 2DEG at the interface between $SrTiO_3$ and $LaAlO_3$ is gate tunable and up to $5 \times 10^{-12}$ eVm, based on previous weak antilocalization measurement at 1.5 K (*27*).

The spin injection experiment is performed via spin pumping, which is a widely used technique to probe the spin-to-charge conversion in nonmagnetic materials (*4, 10, 17, 28-32*). When the ferromagnetic resonance condition for Py is fulfilled under the RF microwave field, a spin current is injected into the Rashba-split 2DEG due to the angular momentum conservation rule. The IEE of the spin current gives rise to an electric field that is in-plane and perpendicular to the spin polarization direction, as shown in Fig. 1D. This electric field could be detected by measuring the voltage on the two ends of the 2DEG at the interface of $SrTiO_3/LaAlO_3$ (see Methods for details).

Fig. 2A shows a typical ferromagnetic resonance (FMR) spectrum of Py on the sample of $SrTiO_3$/6 UC $LaAlO_3$ under the RF frequency of 6 GHz (see Methods for details), where $S_{21}$ is the the forward amplitude of the complex transmission coefficients. The magnetization dynamics of the Py follows the Landau-Lifshitz-Gilbert equation (*33, 34*):

$$\frac{d\vec{M}}{dt} = -\gamma \vec{M} \times \vec{H}_{eff} + \frac{\alpha}{M_S} \vec{M} \times \frac{d\vec{M}}{dt} \quad (2)$$

where $\vec{M}$ is the magnetization vector, $\vec{H}_{eff}$ is the total effective magnetic field, $\gamma$ is the gyromagnetic ratio, $M_S = |\vec{M}|$ is the saturation magnetization, and α is the Gilbert damping constant. At the resonance magnetic field ($H_{res}$) of ~ 470 Oe, the precessing magnetization of Py electrode absorbs the microwave, and as a result, the measured amplitude of the complex transmission coefficient shows a minimum.

The black circles in Fig. 2B correspond to the measured voltage on the sample of $SrTiO_3$/6 UC $LaAlO_3$ as a function of the magnetic field. It is clearly shown that the voltage signal is observed at the magnetic field around the resonance field of the Py on the sample of $SrTiO_3$/6 UC $LaAlO_3$ (Fig. 2A), indicating the measured voltage has a relationship of the spin pumping from Py under the ferromagnetic resonance condition. We can analyze the measured voltage in terms of two major contributions (*4, 16*): the IEE ($V_{IEE}$), due to IEE of the spin polarization in the Rashba-split 2DEG, and the anomalous Hall effect of Py ($V_{AHE}$). The $V_{IEE}$ and $V_{AHE}$ are expected to show different symmetries around $H_{res}$, in which $V_{IEE}$ shows a symmetric Lorentzian shape while $V_{AHE}$ exhibits an anti-symmetric Lorentzian shape. A minor contribution is the Seebeck effect ($V_{SE}$), of which the sign does not depend on the Py magnetization direction (*16*). Thus, we can obtain all these three contributions, $V_{IEE}$, $V_{AHE}$, and $V_{SE}$, based on their different symmetries as a function of the magnetic field. First, the measured voltage is numerically simulated following the equation below:

$$V(H) = V_S \frac{(\Delta H)^2}{(H - H_{res})^2 + (\Delta H)^2} + V_{AHE} \frac{-2\Delta H (H - H_{res})}{(H - H_{res})^2 + (\Delta H)^2} \quad (3)$$

where $V_S$ is the voltage amplitude for the symmetric Lorentzian shape, and $\Delta H$ is the half-line width. Based on the values for $V_S(+H)$ and $V_S(-H)$ obtained for the positive and negative $H$, we can determine the $V_{IEE}$ based on $V_{IEE} = [V_s(+H) - V_s(-H)]/2$ , and $V_{SE}$ based on $V_{SE} = [V_s(+H) + V_s(-H)]/2$ . The red, blue, and green solid lines correspond to the numerical fitted components due to the spin pumping and inverse Edelstein effect, the Seebeck effect, and the anomalous Hall effect, respectively. The resonance frequency ($f_{res}$) vs. $H_{res}$ is shown in Fig. 2C and $\Delta H$ vs. $f_{res}$ for 20 nm Py on 6 UC $LaAlO_3$ is plotted in Fig. 2D. Based on these results, the demagnetization field of 20 nm Py ($4\pi M_{eff}$) is obtained to be 9.1 × $10^3$ Oe using the Kittel formula:

$$f_{res} = (\frac{\gamma}{2\pi})[H_{res}(H_{res} + 4\pi M_{eff})]^{1/2} \quad (4)$$

The Gilbert damping constant of the 20 nm Py on 6 UC $LaAlO_3$ is calculated to be 0.0097 obtained from slope of the linearly fitted curve (red line), which is significantly higher compared to that of the 20 nm Py (0.0064) grown on $SiO_2$ substrate (black circles) (*35*). One major cause for the enhanced Gilbert damping parameter is the spin pumping into the Rashba-split 2DEG from the Py. Based on the enhanced Gilbert damping constants, we have estimated the spin mixing conductance ($G_{\uparrow\downarrow}$) between the magnetization of Py and the spins in Rashba-split 2DEG to be $3.7 \times 10^{19}$ m$^2$ and the injected spin current to be $6.7 \times 10^6$ A/m$^2$ using the model well established for spin pumping(*28, 29, 36*). The spin current value is in the range of previous reported values for the Rashba interface (*4, 37*).

Next, the temperature dependence of the IEE is studied. Fig. 3A shows the representative voltages measured on the $SrTiO_3$/6 UC $LaAlO_3$ at various temperatures from 300 to 100 K. The $V_{IEE}$ decreases quickly as the temperature decreases, and below 200 K, $V_{IEE}$ vanishes and is no longer able to be detected. The temperature dependence of the $V_{IEE}$ on 2DEG between $SrTiO_3$ and 6 UC $LaAlO_3$ is summarized in Fig. 3B. As the temperature decreases, the 2DEG resistance shows a metallic behavior (Fig. 3C), measured by two-probe method using the same Al wires for the IEE voltage measurements, which is consistent with previous studies (*19, 20*). And the junction resistance increases as the temperature decreases (Fig. 3C). The increase of the junction resistance might give rise a lower value of the spin current across $LaAlO_3$ layer, and consequently a lower $V_{IEE}$.

To further study the temperature dependence behavior, we also measure two other samples of $SrTiO_3$/20 UC $LaAlO_3$ and $SrTiO_3$/40 UC $LaAlO_3$. The $V_{IEE}$ decrease quickly as the temperature increases, as shown in Figs. 4A and 4B, which is very similar to that of $SrTiO_3$/6 UC $LaAlO_3$. And $R_J$ increases as temperature decreases, as shown in the inset of Figs. 4A and 4B. Fig. 4C shows the normalized $V^*_{IEE}$ by the RF microwave power ($V^*_{IEE} = V_{IEE}/P_{RF}$) as a function of the $LaAlO_3$ thickness. As the thickness increases, the $V^*_{IEE}$ decreases.

The physical properties of the 2DEG between $SrTiO_3$ and $LaAlO_3$ could be largely modulated by a perpendicular electric field (*25, 27, 38*). To study the $V_{IEE}$ as a function of the gate voltage at room temperature, the heterostructures consisting of $SrTiO_3$ and 3 UC $LaAlO_3$ is chosen due to the large tunability of the electrical properties of the Rashba-split 2DEG at the interface. An electrode of silver paste is used on the other side of $SrTiO_3$ substrate to serve as a back gate. The schematic of the measurement is illustrated in the inset of Fig. 5A, and the Fig. 5A shows the spin voltage measured as a function of magnetic field with the RF power of 0.45 W under the gate voltage of -20, -8, 0, 100, and 200 V, respectively. Clearly, no spin signal is observable under the gate voltage of -20 V, while clear spin signals are observed under the gate voltage ($V_G$) between 0 and 200 V. Fig. 5B summarizes the spin voltage as a function of the back gate voltage measured at 300 K. Then we check the resistance of the 2DEG at the interface as a function of the gate voltage. It is noted that the resistance of the 2DEG greatly increases as the gate voltage goes to be negative, as shown in Fig. 5C. The spin pumping, angular momentum transfer from Py to the spin polarization in the 2DEG, via such thick LAO layer is very interesting. To our best understanding, the angular momentum could be transferred from Py layer to spins in the 2DEG across this insulating LAO layer via two possible mechanism: spin tunneling across the LAO layer and the angular momentum transfer via defects in the LAO layer (i.e. oxygen vacancies).

## Discussion

The temperature dependence of the $V_{IEE}$ needs further theoretical and experimental studies. There is no existing mechanisms that could fully explain our observation, as discussed in following. Firstly of all, it is noted that there is no direct association of this behavior with the junction resistance, which affect the spin pumping efficiency. As shown in Figs. 4A and 4B, although $R_J$ varies over several orders of magnitude for various 2DEG between $SrTiO_3$ and $LaAlO_3$ of 20, and 40 UCs, the magnitudes of $V_{IEE}$ for these two samples exhibit little difference, indicating that the IEE signal trend cannot simply be explained by the $R_J$ variation. Especially, the disappearance of the $V_{IEE}$ happens below a critical temperature of ~ 200 K in all three samples where $R_J$ is very different. Second of all, we measure the temperature and thickness dependence of the mobility and carrier density of the interface, which are expected to affect the Rashba spin orbit coupling and thus the spin to charge conversion efficiency. As shown in Fig. S1, the temperature and thickness dependences of the mobility and carrier density show distinct trends compared to those of the $V_{IEE}$, indicating no obvious relationship between the carrier densities, the mobility at the interface and IEE signal. Thirdly of all, if there is strong magnetic impurity scattering at the interface, which could provide strong spin scattering to destroy the spin-momentum locking, and make the $V_{IEE}$ signal disappear. However, if the magnetic impurities exist, the spin scattering is expected to happen at all temperatures, especially at high temperatures. This is not likely the mechanism that could account for our results. Fourthly of all, the exchange interactions between the Py and the interfacial 2DEG of $SrTiO_3/LaAlO_3$ might play an important role. If the spin injection and accumulation mechanisms are due to the exchange coupling between the Py and 2DEG (*37*), the exchange interaction could be strongly temperature dependent. However, exchange interaction is usually an energy scale that is determined by the overlap of wave-functions and temperature independent. There is no cause for that the exchange interactions disappears significantly below ~ 200 K, if temperature dependent exchange interaction is the mechanism. Lastly, the observation might be related to the spin transport via the acoustic phonons in the $LaAlO_3$ layer, where the acoustic phonons could be chiral based on theoretical calculations for $LaAlO_3$ of four-fold square symmetry (*39*). However, there is no results indicating crystal structure change yet.

At room temperature, the gate voltage provides a powerful tool to tune the spin-to-charge conversion efficiency and even to turn on/off the signal of $V_{IEE}$ (Fig. 5 and Fig. S2). The $V_{IEE}$, corresponding to the effective charge current, generated at negative gate voltages, is significantly lower, which means a lower effective spin-to-charge conversion. Whilst, the large spin signal and low resistance of the 2DEG under positive gate voltage indicate a larger spin-to-charge conversion.

As it is known that the dielectric constant of $SrTiO_3$ increases significantly at low temperatures (*40*), the gate voltage modulation of the $V_{IEE}$ at low temperatures is expected to be significantly enhanced.

In summary, we have investigated the IEE in the Rashba-split 2DEG formed between two insulating oxides $SrTiO_3$ and $LaAlO_3$ and demonstrated the gate voltage modulation of the $V_{IEE}$ at room temperature. Our results reveal that the oxide interface can be used for efficient charge-to-spin conversion for the generation and detection of spin current beyond the applications for spin channels (*41-43*) and the future electronics (*19-23*).

*Note added.* During the preparation of this manuscript, we became aware of related spin pumping and IEE studies in $SrTiO_3$/***2 UC*** $LaAlO_3$ at low temperature (T = ***7 K***) by E. Lesne *et al* (*37*). Different from Ref. 37, we report the strong modulation of IEE at room temperature by an electric field, the strong temperature dependence of the IEE, and the spin injection into the Rashba-split 2DEG across a series of LAO thicknesses up to 40 UC.

## Materials and Methods

### Materials growth

The $LaAlO_3$ layers from 3 to 40 UCs were grown on the top of (001)-oriented $SrTiO_3$ substrates ($5 \times 5$ mm$^2$) via pulsed laser deposition with the laser fluence of 0.7 Jcm$^{-2}$ and the repetition rate of 1 Hz. During the growth, the $SrTiO_3$ substrates were held at the temperature of 800 °C and in the oxygen pressure of $1\times10^{-5}$ mbar. *In situ* RHEED oscillations was used to monitor the thickness of the $LaAlO_3$. After deposition, the samples were *in situ* annealed at 600 °C for 1 h and then cooled back to room temperature in 200 mbar of $O_2$. The details of the growth could be found in our earlier report (*44*).

### Device fabrication

The device structure for the IEE measurement was illustrated in Fig. 1D. The 20 nm Py was deposited onto the $LaAlO_3$ via a shadow mask technique (size: ~ $3 \times 4$ mm$^2$) using RF magnetron sputtering and a 3 nm Al was deposited *in situ* to prevent the oxidation of Py. The voltages were detected on the two ends of the 2DEG via Al bonding wires.

### IEE measurement

The IEE response was measured using a digital lock-in amplifier (SRS Inc. SR830) by modulating the microwave FR power at the frequency of 17 Hz for a better signal-to-noise ratio. The FMR measurement of Py magnetization dynamics was performed using a vector network analyzer (VNA, Agilent E5071C). For the IEE measurement, the RF microwave was applied at a frequency of 6 GHz and a RF power of 1.25 W unless noted otherwise. Both FMR and IEE measurements were performed using the coplanar waveguide technique in the variable temperature insert of a Quantum Design Physical Properties Measurement System (PPMS).

**H2: Supplementary Materials**

Fig. S1. The electron transport properties of the 2DEG.

Fig. S2. The gate voltage dependence of IEE of the Rashba-split 2DEG between $SrTiO_3$ and 3 UC $LaAlO_3$.

**References and Notes:**

1. V. M. Edelstein, Spin polarization of conduction electrons induced by electric current in two-dimensional asymmetric electron systems. *Solid State Commun.* **73**, 233-235 (1990).
2. A. Manchon, H. C. Koo, J. Nitta, S. M. Frolov, R. A. Duine, New perspectives for Rashba spin-orbit coupling. *Nat. Mater.* **14**, 871-882 (2015).
3. K. Shen, G. Vignale, R. Raimondi, Microscopic Theory of the Inverse Edelstein Effect. *Phys. Rev. Lett.* **112**, 096601 (2014).
4. J. C. R. Sánchez, L. Vila, G. Desfonds, S. Gambarelli, J. P. Attané, J. M. De Teresa, C. Magén, A. Fert, Spin-to-charge conversion using Rashba coupling at the interface between non-magnetic materials. *Nat. Commun.* **4**, (2013).
5. H. J. Zhang, S. Yamamoto, B. Gu, H. Li, M. Maekawa, Y. Fukaya, A. Kawasuso, Charge-to-Spin Conversion and Spin Diffusion in Bi/Ag Bilayers Observed by Spin-Polarized Positron Beam. *Phys. Rev. Lett.* **114**, 166602 (2015).
6. W. Zhang, M. B. Jungfleisch, W. Jiang, J. E. Pearson, A. Hoffmann, Spin pumping and inverse Rashba-Edelstein effect in NiFe/Ag/Bi and NiFe/Ag/Sb. *J. Appl. Phys.* **117**, 17C727 (2015).
7. M. Isasa, M. C. Martínez-Velarte, E. Villamor, C. Magén, L. Morellón, J. M. De Teresa, M. R. Ibarra, G. Vignale, E. V. Chulkov, E. E. Krasovskii, L. E. Hueso, F. Casanova, Origin of inverse Rashba-Edelstein effect detected at the Cu/Bi interface using lateral spin valves. *Phys. Rev. B* **93**, 014420 (2016).
8. J. B. S. Mendes, O. Alves Santos, L. M. Meireles, R. G. Lacerda, L. H. Vilela-Leão, F. L. A. Machado, R. L. Rodríguez-Suárez, A. Azevedo, S. M. Rezende, Spin-Current to Charge-Current Conversion and Magnetoresistance in a Hybrid Structure of Graphene and Yttrium Iron Garnet. *Phys. Rev. Lett.* **115**, 226601 (2015).
9. W. Han, Perspectives for spintronics in 2D materials. *APL Mater.* **4**, 032401 (2016).
10. S. Dushenko, H. Ago, K. Kawahara, T. Tsuda, S. Kuwabata, T. Takenobu, T. Shinjo, Y. Ando, M. Shiraishi, Gate-Tunable Spin-Charge Conversion and the Role of Spin-Orbit Interaction in Graphene. *Phys. Rev. Lett.* **116**, 166102 (2016).
11. W. Zhang, J. Sklenar, B. Hsu, W. Jiang, M. B. Jungfleisch, J. Xiao, F. Y. Fradin, Y. Liu, J. E. Pearson, J. B. Ketterson, Z. Yang, A. Hoffmann, Research Update: Spin transfer torques in permalloy on monolayer MoS2. *APL Mater.* **4**, 032302 (2016).
12. D. MacNeill, G. M. Stiehl, M. H. D. Guimaraes, R. A. Buhrman, J. Park, D. C. Ralph, Control of spin-orbit torques through crystal symmetry in WTe2/ferromagnet bilayers. *Nat. Phys.* **advance online publication**, (2016).
13. A. R. Mellnik, J. S. Lee, A. Richardella, J. L. Grab, P. J. Mintun, M. H. Fischer, A. Vaezi, A. Manchon, E. A. Kim, N. Samarth, D. C. Ralph, Spin-transfer torque generated by a topological insulator. *Nature* **511**, 449-451 (2014).

14. C. H. Li, O. M. J. van`t Erve, J. T. Robinson, Y. Liu, L. Li, B. T. Jonker, Electrical detection of charge-current-induced spin polarization due to spin-momentum locking in Bi2Se3. *Nat. Nanotechnol.* **9**, 218-224 (2014).
15. Y. Fan, P. Upadhyaya, X. Kou, M. Lang, S. Takei, Z. Wang, J. Tang, L. He, L.-T. Chang, M. Montazeri, G. Yu, W. Jiang, T. Nie, R. N. Schwartz, Y. Tserkovnyak, K. L. Wang, Magnetization switching through giant spin–orbit torque in a magnetically doped topological insulator heterostructure. *Nat. Mater.* **13**, 699-704 (2014).
16. Y. Shiomi, K. Nomura, Y. Kajiwara, K. Eto, M. Novak, K. Segawa, Y. Ando, E. Saitoh, Spin-Electricity Conversion Induced by Spin Injection into Topological Insulators. *Phys. Rev. Lett.* **113**, 196601 (2014).
17. J. C. Rojas-Sánchez, S. Oyarzún, Y. Fu, A. Marty, C. Vergnaud, S. Gambarelli, L. Vila, M. Jamet, Y. Ohtsubo, A. Taleb-Ibrahimi, P. Le Fèvre, F. Bertran, N. Reyren, J. M. George, A. Fert, Spin to Charge Conversion at Room Temperature by Spin Pumping into a New Type of Topological Insulator: $\ensuremath{\alpha}$-Sn Films. *Phys. Rev. Lett.* **116**, 096602 (2016).
18. Q. Song, J. Mi, D. Zhao, T. Su, W. Yuan, W. Xing, Y. Chen, T. Wang, T. Wu, X. H. Chen, X. C. Xie, C. Zhang, J. Shi, W. Han, Spin injection and inverse Edelstein effect in the surface states of topological Kondo insulator SmB6. *Nat. Commun.* **7**, 13485 (2016).
19. H. Y. Hwang, Y. Iwasa, M. Kawasaki, B. Keimer, N. Nagaosa, Y. Tokura, Emergent phenomena at oxide interfaces. *Nat Mater* **11**, 103-113 (2012).
20. J. Mannhart, D. G. Schlom, Oxide Interfaces—An Opportunity for Electronics. *Science* **327**, 1607-1611 (2010).
21. A. Ohtomo, H. Y. Hwang, A high-mobility electron gas at the LaAlO3/SrTiO3 heterointerface. *Nature* **427**, 423-426 (2004).
22. N. Reyren, S. Thiel, A. D. Caviglia, L. F. Kourkoutis, G. Hammerl, C. Richter, C. W. Schneider, T. Kopp, A.-S. Rüetschi, D. Jaccard, M. Gabay, D. A. Muller, J.-M. Triscone, J. Mannhart, Superconducting Interfaces Between Insulating Oxides. *Science* **317**, 1196-1199 (2007).
23. S. Stemmer, S. James Allen, Two-Dimensional Electron Gases at Complex Oxide Interfaces. *Annu. Rev. Mater. Res.* **44**, 151-171 (2014).
24. M. Basletic, J. L. Maurice, C. Carretero, G. Herranz, O. Copie, M. Bibes, E. Jacquet, K. Bouzehouane, S. Fusil, A. Barthelemy, Mapping the spatial distribution of charge carriers in LaAlO3/SrTiO3 heterostructures. *Nat. Mater.* **7**, 621-625 (2008).
25. C. Cen, S. Thiel, G. Hammerl, C. W. Schneider, K. E. Andersen, C. S. Hellberg, J. Mannhart, J. Levy, Nanoscale control of an interfacial metal-insulator transition at room temperature. *Nat. Mater.* **7**, 298-302 (2008).
26. S. Hong, V. Diep, S. Datta, Y. P. Chen, Modeling potentiometric measurements in topological insulators including parallel channels. *Phys. Rev. B* **86**, 085131 (2012).
27. A. D. Caviglia, M. Gabay, S. Gariglio, N. Reyren, C. Cancellieri, J. M. Triscone, Tunable Rashba Spin-Orbit Interaction at Oxide Interfaces. *Phys. Rev. Lett.* **104**, 126803 (2010).
28. O. Mosendz, J. E. Pearson, F. Y. Fradin, G. E. W. Bauer, S. D. Bader, A. Hoffmann, Quantifying Spin Hall Angles from Spin Pumping: Experiments and Theory. *Phys. Rev. Lett.* **104**, 046601 (2010).
29. K. Ando, S. Takahashi, J. Ieda, Y. Kajiwara, H. Nakayama, T. Yoshino, K. Harii, Y. Fujikawa, M. Matsuo, S. Maekawa, E. Saitoh, Inverse spin-Hall effect induced by spin pumping in metallic system. *J. Appl. Phys.* **109**, 103913 (2011).

30. K. Ando, S. Takahashi, J. Ieda, H. Kurebayashi, T. Trypiniotis, C. H. W. Barnes, S. Maekawa, E. Saitoh, Electrically tunable spin injector free from the impedance mismatch problem. *Nat Mater* **10**, 655-659 (2011).
31. F. D. Czeschka, L. Dreher, M. S. Brandt, M. Weiler, M. Althammer, I. M. Imort, G. Reiss, A. Thomas, W. Schoch, W. Limmer, H. Huebl, R. Gross, S. T. B. Goennenwein, Scaling Behavior of the Spin Pumping Effect in Ferromagnet-Platinum Bilayers. *Phys. Rev. Lett.* **107**, 046601 (2011).
32. K. Ando, E. Saitoh, Observation of the inverse spin Hall effect in silicon. *Nat Commun* **3**, 629 (2012).
33. L. Landau, E. Lifshitz, On the theory of the dispersion of magnetic permeability in ferromagnetic bodies. *Phys. Z. Sowjetunion* **8**, 153 (1935).
34. T. L. Gilbert, A phenomenological theory of damping in ferromagnetic materials. *Magnetics, IEEE Transactions on* **40**, 3443-3449 (2004).
35. Y. Zhao, Q. Song, S.-H. Yang, T. Su, W. Yuan, S. S. P. Parkin, J. Shi, W. Han, Experimental Investigation of Temperature-Dependent Gilbert Damping in Permalloy Thin Films. *Scientific Reports* **6**, 22890 (2016).
36. Y. Tserkovnyak, A. Brataas, G. E. W. Bauer, B. I. Halperin, Nonlocal magnetization dynamics in ferromagnetic heterostructures. *Rev. Mod. Phys.* **77**, 1375-1421 (2005).
37. E. Lesne, Y. Fu, S. Oyarzun, J. C. Rojas-Sanchez, D. C. Vaz, H. Naganuma, G. Sicoli, J. P. Attane, M. Jamet, E. Jacquet, J. M. George, A. Barthelemy, H. Jaffres, A. Fert, M. Bibes, L. Vila, Highly efficient and tunable spin-to-charge conversion through Rashba coupling at oxide interfaces. *Nat. Mater.* **15**, 1261-1266 (2016).
38. S. Thiel, G. Hammerl, A. Schmehl, C. W. Schneider, J. Mannhart, Tunable Quasi-Two-Dimensional Electron Gases in Oxide Heterostructures. *Science* **313**, 1942-1945 (2006).
39. L. Zhang, Q. Niu, Chiral Phonons at High-Symmetry Points in Monolayer Hexagonal Lattices. *Phys. Rev. Lett.* **115**, 115502 (2015).
40. K. A. Müller, H. Burkard, $SrTiO_3$: An intrinsic quantum paraelectric below 4 K. *Phys. Rev. B* **19**, 3593-3602 (1979).
41. N. Reyren, M. Bibes, E. Lesne, J. M. George, C. Deranlot, S. Collin, A. Barthélémy, H. Jaffrès, Gate-Controlled Spin Injection at $LaAlO_3/SrTiO_3$ Interfaces. *Phys. Rev. Lett.* **108**, 186802 (2012).
42. W. Han, X. Jiang, A. Kajdos, S.-H. Yang, S. Stemmer, S. S. P. Parkin, Spin injection and detection in lanthanum- and niobium-doped SrTiO3 using the Hanle technique. *Nat. Commun.* **4**, (2013).
43. R. Ohshima, Y. Ando, K. Matsuzaki, T. Susaki, M. Weiler, S. Klingler, H. Huebl, E. Shikoh, T. Shinjo, S. T. B. Goennenwein, M. Shiraishi, Realization of d-electron spin transport at room temperature at a LaAlO3/SrTiO3 interface. *arXiv:1601.07568v1*, (2016).
44. Y. Lei, Y. Li, Y. Z. Chen, Y. W. Xie, Y. S. Chen, S. H. Wang, J. Wang, B. G. Shen, N. Pryds, H. Y. Hwang, J. R. Sun, Visible-light-enhanced gating effect at the LaAlO3/SrTiO3 interface. *Nat. Commun.* **5**, (2014).

## Acknowledgments

**General**: We acknowledge the fruitful discussion with Qian Niu, Xincheng Xie and Lifa Zhang. Financial support from National Basic Research Programs of China (973 program

Grant Nos. 2015CB921104, 2016YFA0300701, 2014CB920902, and 2013CB921901) and National Natural Science Foundation of China (NSFC Grant No. 11574006 and 11520101002) is greatly acknowledged.

**Author contributions:** J.S., J.R.S., and W.H. proposed and surprised the studies. Q.S. did the device fabrication and the IEE measurements. Q.S. and H.Z. analyzed the data. H.Z. and J.R.S. grew the $SrTiO_3$/$LaAlO_3$ samples and performed the characterization of the electrical properties of the 2DEG. Q.S., J.S., J.R.S, and W.H. wrote the manuscript. All authors commented on the manuscript and contributed to its final version.

**Competing interests:** The authors declare no competing interests.

**Data Availability:** All data needed to evaluate the conclusions in the paper are present in the paper and/or the Supplementary Materials. Additional data related to this paper may be requested from the authors.

**Figures and Tables**

# Figure 1

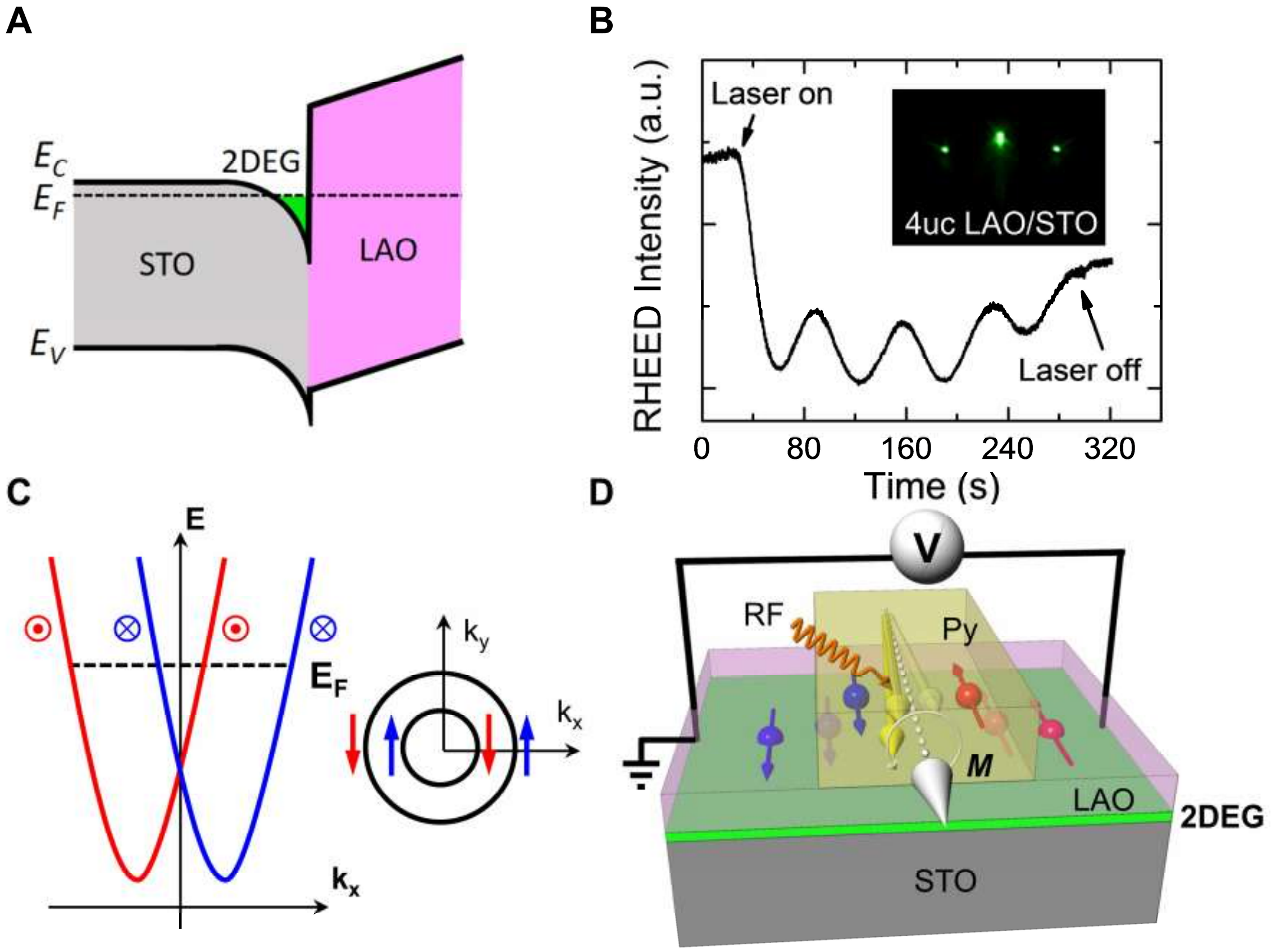

**Fig. 1. The Rashba-split 2DEG between $SrTiO_3$ and $LaAlO_3$. A,** Schematic drawing of 2DEG and the band alignment for the $SrTiO_3$ and $LaAlO_3$ heterostructures. **B,** The RHEED oscillations of 4 UC $LaAlO_3$ growing on $SrTiO_3$. **C,** The energy dispersion for a typical Rashba spin-split 2DEG. At the Fermi level, the outer-circle and inner-circles exhibit the opposite spin textures. **D,** Schematic drawing of the IEE measurements. The spin current is injected via spin pumping from the Py under ferromagnetic resonance. A voltage meter is used to probe the electric field generated due to the IEE of the Rashba-split 2DEG between two insulating oxides, $SrTiO_3$ and $LaAlO_3$.

# Figure 2

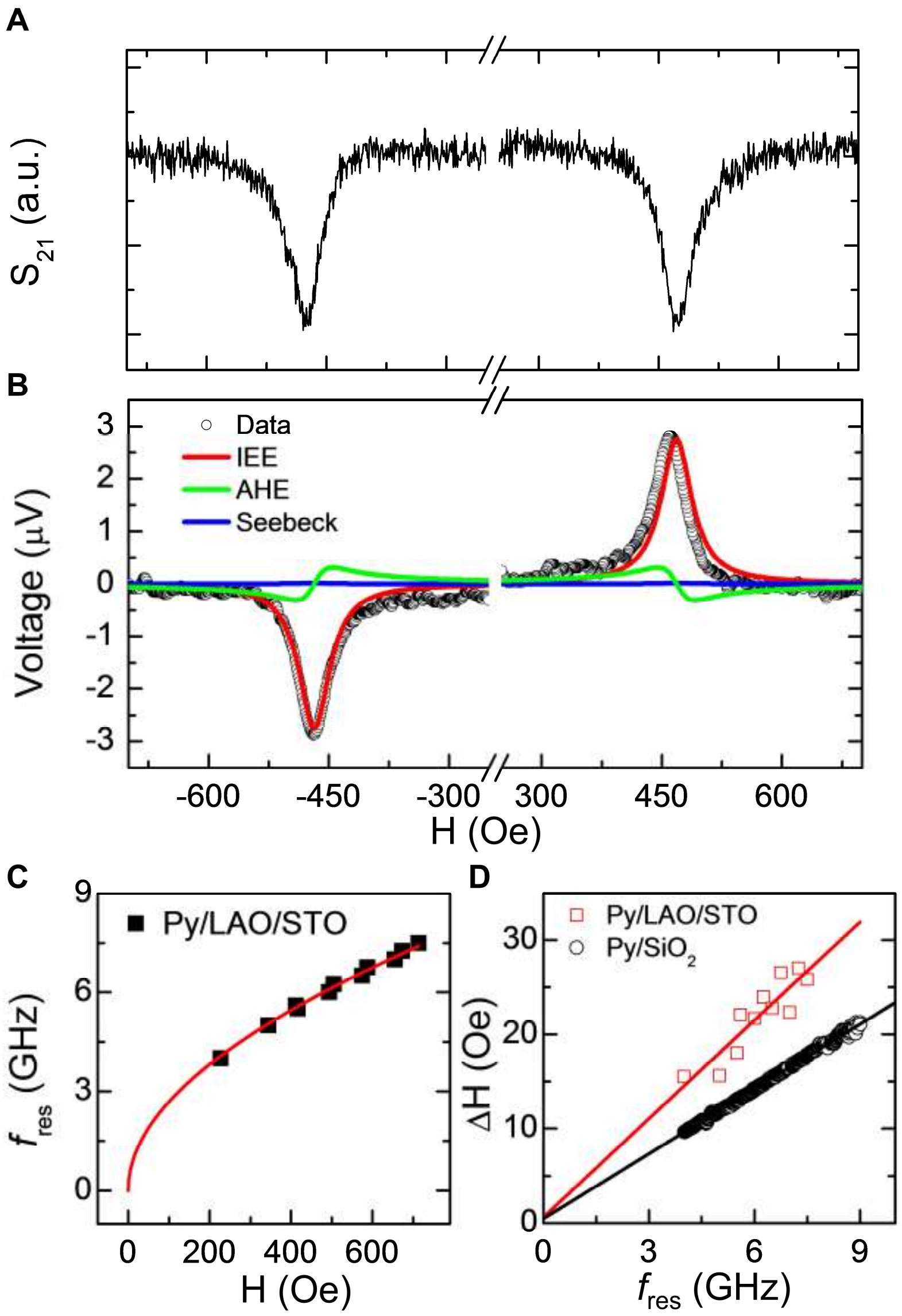

**Fig. 2. The electrical detection of IEE of the Rashba-split 2DEG between $SrTiO_3$ and 6 UC $LaAlO_3$ at 300 K. A,** Representative ferromagnetic resonance spectra of the Py electrode on the $SrTiO_3$/6 UC $LaAlO_3$ using a vector network analyzer with RF frequency of 6 GHz. **B,** The measured voltage (black circles) as a function of the magnetic field using a signal generator with the power of 1.25 W and RF frequency of 6 GHz. The red lines, blue lines and green lines are fitted curves that correspond to the voltages due to IEE of the injected spin current, the Seebeck effect, and the anomalous Hall effect of the Py, respectively. **C,** The resonance frequency ($f_{res}$) as a function of the resonance magnetic field ($H_{res}$). The solid line is a fitted curve based on the Kittel formula. **D,** The half linewidth (ΔH) vs. the resonance frequency for Py on 6 UC $LaAlO_3$ (red squares) and $SiO_2$ (black circles) at 300 K, from which the Gilbert damping can be obtained from the slope of the linearly fitted curves.

# Figure 3

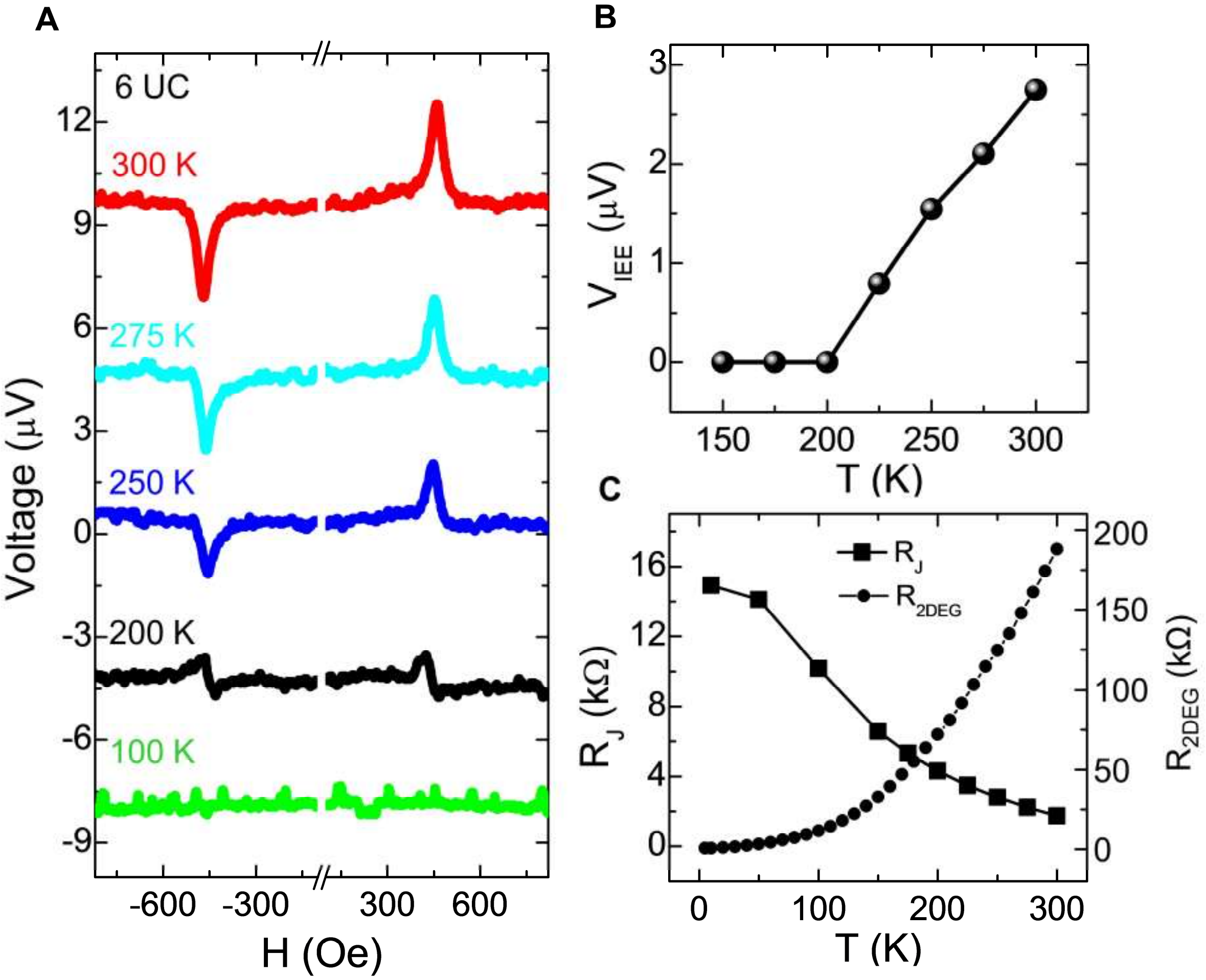

**Fig. 3. The temperature dependence of IEE of the Rashba-split 2DEG between $SrTiO_3$ and 6 UC $LaAlO_3$. A,** The measured voltage (black circles) on $SrTiO_3$/6 UC $LaAlO_3$ as a function of the magnetic field at 300, 275, 250, 200, and 100 K, respectively. **B,** The temperature dependence of $V_{IEE}$ of the Rashba-split 2DEG between $SrTiO_3$/6 UC $LaAlO_3$. **C,** The temperature dependence of the junction resistance ($R_J$) between the Py and the 2DEG resistance ($R_{2DEG}$) between $SrTiO_3$ and 6 UC $LaAlO_3$.

# Figure 4

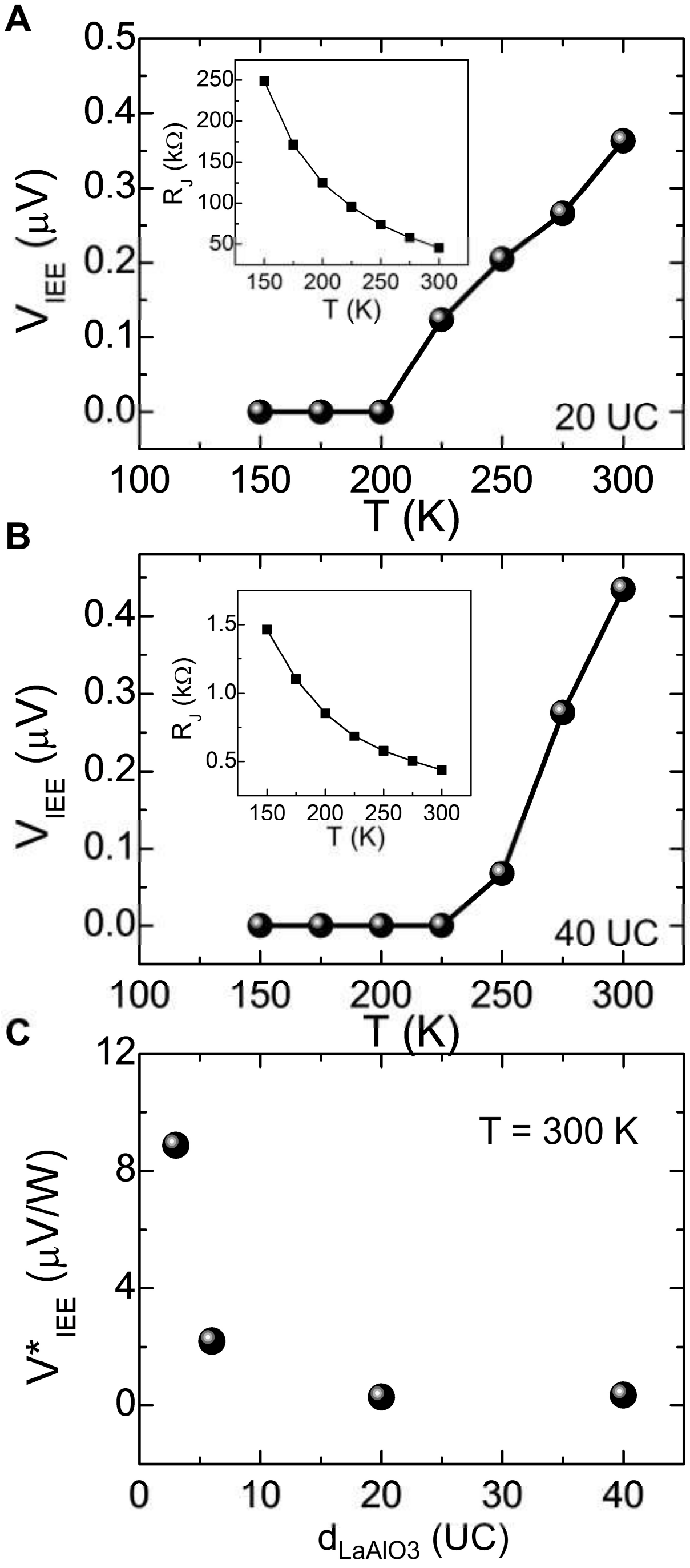

**Fig. 4. The temperature and $LaAlO_3$ thickness dependences of IEE for the Rashba-split 2DEGs for $SrTiO_3/LaAlO_3$ with thickness up to 40 UCs. A-B,** The temperature dependence of $V_{IEE}$ for the Rashba-split 2DEGs for $SrTiO_3$/20 UC $LaAlO_3$ and $SrTiO_3$/40 UC $LaAlO_3$, respectively. Inset: the temperature dependence of $R_J$ between Py and the 2DEG between $SrTiO_3$ and $LaAlO_3$. **C,** The normalized $V^*_{IEE}$ as a function of the $LaAlO_3$ thickness.

# Figure 5

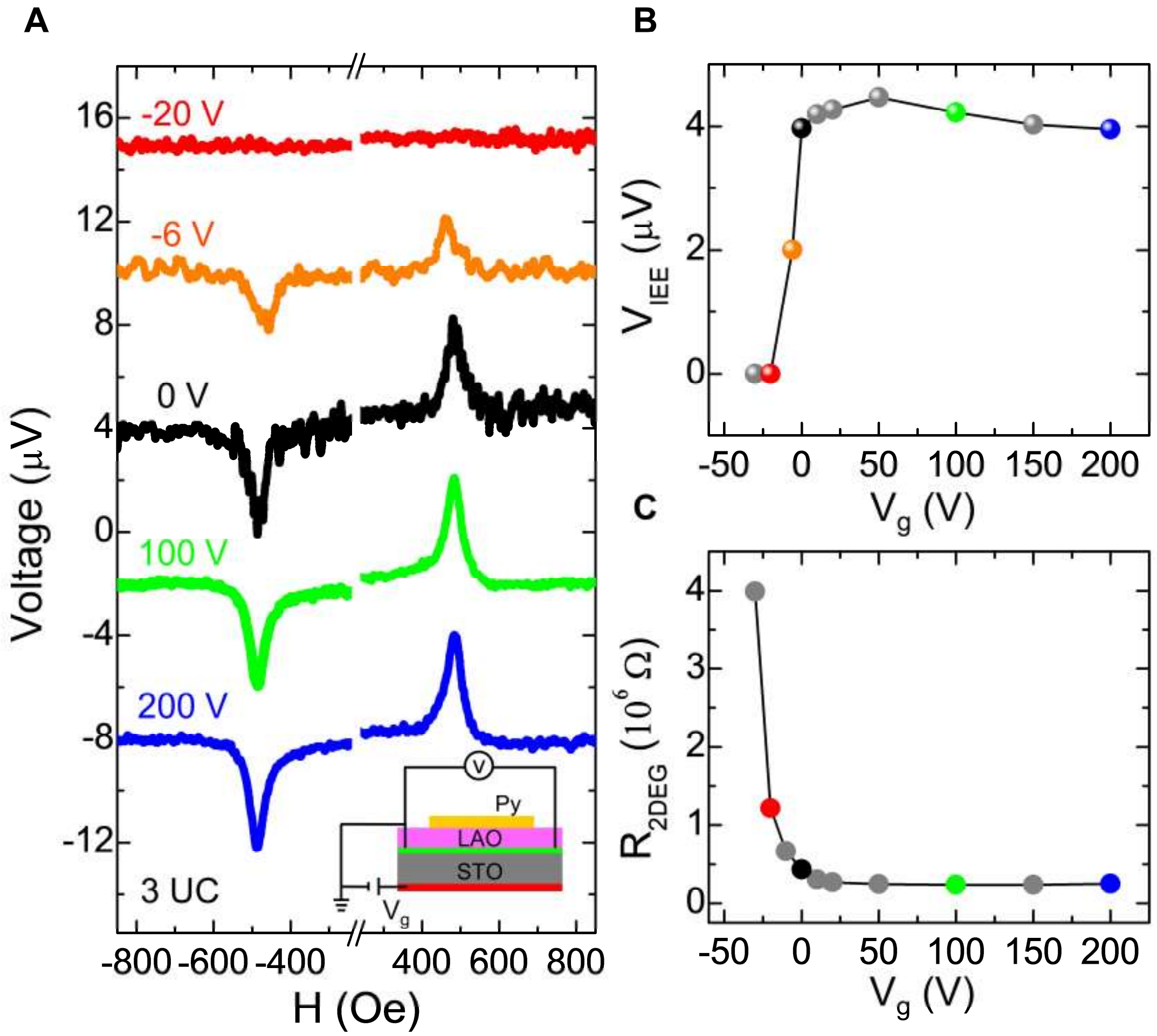

**Fig. 5. The gate voltage dependence of IEE of the Rashba-split 2DEG between $SrTiO_3$ and 3 UC $LaAlO_3$. A,** The measured voltage on $SrTiO_3$/3 UC $LaAlO_3$ as a function of the magnetic field at 300 K for Vg = -20, -6, 0, 100, and 200 V, respectively. Inset: Schematic drawing of the measurement under electric field using the $SrTiO_3$ as the dielectric layer. **B,** The gate voltage dependence of the $V_{IEE}$ for the Rashba-split 2DEG between $SrTiO_3$/3 UC $LaAlO_3$. **C,** The gate voltage dependence of 2DEG resistance between SrTiO3/3 UC $LaAlO_3$.